\documentclass[11pt]{article}
\usepackage{moriond,epsfig}

\def\Journal#1#2#3#4{{#1} {\bf #2}, #3 (#4)}

\def\NPA{{\em Nucl. Phys.} A}

\def\PLB{{\em Phys. Lett.}  B}
\def\PRL{\em Phys. Rev. Lett.}
\def\PRC{{\em Phys. Rev.} C}

\def\be{\begin{equation}}
\def\ee{\end{equation}}
\def\bea{\begin{eqnarray}}
\def\eea{\end{eqnarray}}

\begin{document}
\vspace*{4cm}
\title{PIONIC HYDROGEN AT PSI}

\author{ D.GOTTA~\footnote{for the PIONIC HYDROGEN collaboration}}

\address{Institut f\"ur Kernphysik, Forschungszentrum J\"ulich, D-52425 J\"ulich, Germany\\}

\maketitle\abstracts{
The measurement of the pion--nucleon scattering lengths constitutes a high--precision test of 
the methods of Chiral Pertubation Theory ($\chi$PT), which is the low--energy approach of QCD. 
The $\pi$N s--wave scattering lengths are related to the strong--interaction shift  $\it\epsilon_{1s}$ and 
width $\it\Gamma_{1s}$ of the $s$--states of the pionic hydrogen atom. $\it\epsilon_{1s}$ and 
$\it\Gamma_{1s}$  are determined from the measured energies and line widths of X--ray transitions to the 
$1s$ ground state when compared to the calculated electromagnetic values. A new experiment, set up 
at the  Paul--Scherrer--Institut (PSI), has completed a first series of  measurements.}

\section{Introduction}

The scattering lengths of elastic $a_{\pi^{-}p}\rightarrow a_{\pi^{-}p}$ and charge--exchange 
reaction $a_{\pi^{-}p}\rightarrow a_{\pi^{0}n}$ are related to $\it\epsilon_{1s}$ and 
$\it\Gamma_{1s}$  by Deser--type formulae\,\cite{Des54,Ras82}. Rewritten in terms 
of  two isospin scattering lengths, they describe completely the $\pi  N$ interaction at threshold 
in the isospin symmetric limit. Furthermore, from $\it\Gamma_{1s}$  the $\pi  N$ coupling 
constant $f^{2}_{\pi  N}$ is obtained by the Goldberger--Miyazawa--Oehme sum rule\,\cite{Gol55}. 
Within the frame work of Heavy Baryon $\chi$PT the exotic--atom parameters $\it\epsilon_{1s}$ and 
width $\it\Gamma_{1s}$ and the scattering lengths unambiguously related. In an expansion in momenta, 
the fine structure constant $\alpha $ and quark mass difference, $(m_{d}^{2}-m_{u}^{2})$ in the 
case onf pions,  both electromagnetic and strong interaction are taken into account 
on the same footing\,\cite{Eck95,Bea02}. 

The extracted values for the isospin scattering lengths and for $f^{2}_{\pi  N}$ 
may be compared to the zero order results as given by current algebra\,\cite{Wei66}. 
The diffrences are due to the higher order terms from the chiral expansion
including isopspin--breaking effects not only from the electromagnetic interaction
but also the from the mass difference of the $u$ and $d$ quark. Such contributions may be expressed as 
correction terms $\delta_{\epsilon,\Gamma}$ in the Deser formulae\,\cite{Lyu00}. In addition, 
the pionic--atom results must be consistent with the extrapolation of $\pi$N scattering data to 
threshold and phase--shift analyses\,\cite{menu01}.

Higher--order terms of the chiral expansion contain low--energy constants (LECs) 
to be fixed from data\,\cite{Eck95}. Their uncertainty may be denoted as the theoretical error. 
A recent calculation for $\delta_{\epsilon}$ yields $(-7.2\pm 2.9)\%$\,\cite{Gas03}, the uncertainty 
of which is given mainly by one particular LEC ($f_1$). At present,  
$\delta_{\Gamma}$ is subject of detailed theoretical studies. Here $f_1$ does not appear 
in next--to--leading order, which reduces the uncertainty substantially\,\cite{Rus03}.
Hence, the new experiment\,\cite{R98.01} aims at a significant increase 
in accuracy for $\it\Gamma_{1s}$.

\section{Experimental Approach}

The pionic atom, which is formed in a highly 
excited state, de--excites by X--ray emission and various non--radiative mechanisms. 
The atomic cascade ends in an $s$ state, where a nuclear reaction takes place
(Fig.\,1). One important cascade mechanism is Coulomb 
de--excitation, where the energy release for step $n \rightarrow n'$ is converted 
into kinetic energy of the collision partners $\pi H$ and $H$ (from a molecule $H_2$) 
leading to Doppler broadening of subsequent X--ray transitions. 

\begin{figure}[h]
\begin{center}
\epsfig{file=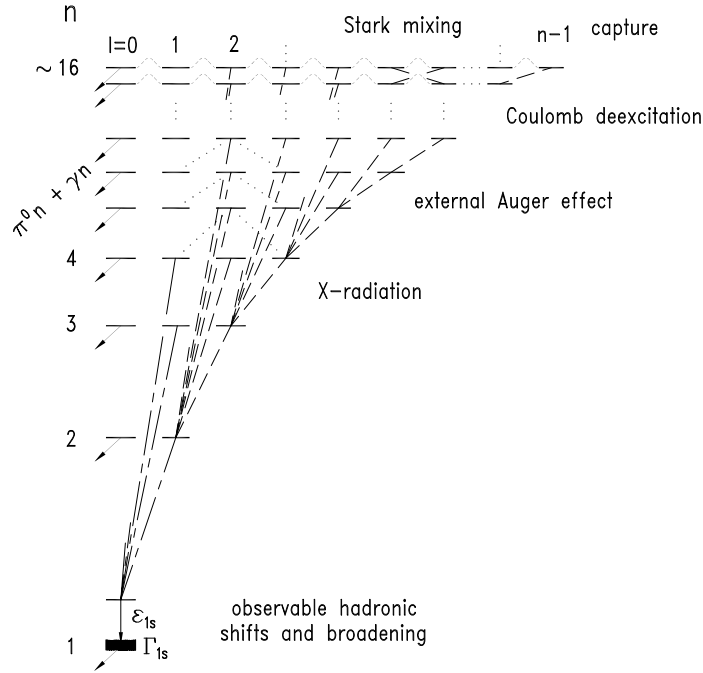,scale=0.48}
\caption{Atomic cascade in pionic hydrogen. The pure electromagnetic energies of
the $2p-1s$, $3p-1s$, and $4p-1s$ transitions are 2.43, 2.88, and 3.04~keV.
The hadronic parameters $\epsilon_{ns}$ and $\Gamma_{ns}$  scale with $1/n^{3}$.}
\end{center}
\label{fig:cascadpih}
\end{figure}

The previous precision experiment -- measuring the $3p-1s$ transitions at
a density equivalent of 15 bar -- yielded $\epsilon_{1s}=+7.108\pm 0.047$~eV
and $\Gamma_{1s}= 0.868\pm 0.078$~eV\,\cite{Sch01}. Whereas the 
precise extraction of the isospin scattering lengths from $\epsilon_{1s}$ is hindered 
at present by the large theoretical error of  $\delta_{\epsilon}$, the uncertainty of 
$\Gamma_{1s}$ is dominated by the correction for the Doppler broadening, because the 
contributions from Coulomb de--excitation are quantitatively not well understood. 

For that reason the precisely measured $1s$--level shift in pionic deuterium was combined 
with $\epsilon_{1s}$ from hydrogen to determine the $\pi N$ scattering lengths and 
$f^{2}_{\pi  N}$ \,\cite{Eri02}. This procedure, however, requires a sophisticated treatment 
of the 3--body system $\pi D$. Furthermore, collisions of the $\pi D$ atom probably 
lead to the formation of complex molecules like [($\pi$dd)d]ee, a process well known 
in the case of muons. Up to now it cannot be excluded that such systems decay with a significant 
fraction by X--ray emission\,\cite{KorWal01}. This leads to small energy shifts hidden by the 
large line width\,\cite{Jon99}.

Both  Coulomb de--excitation and molecule formation are collisional processes and, hence, 
depend on the collision rate, i.~e., on density. Consequently, the strategy of the
new experiment is to measure the density dependence of energy and
line width to identify and/or quantify the above-mentioned cascade effects:
\begin{itemize}
\item
Any contribution to $\epsilon_{1s}$ from the formation of a molecular complex like 
[($\pi$pp)p]ee has to be excluded by establishing a density independent value for the
transition energy.
\item
Information on Coulomb de--excitation is obtained by 
measuring besides the $3p-1s$ also the $4p-1s$ and the $2p-1s$ transitions
and  at least one of these at various densities.
\item
Measure the line widths in muonic hydrogen, where no strong--interaction
broadening occurs, for various transitions and densities.
\end{itemize}
The new experiment is based on techniques developed and applied to 
the precision spectroscopy of X--rays from antiprotonic and pionic 
atoms together with substantial improvements in background suppression\,\cite{Ana01}. 
The cyclotron trap provides a concentrated X--ray source for a focusing 
low--energy crystal spectrometer. X--rays emitted from the stop volume 
are reflected by spherically bent silicon or quartz Bragg crystals of 10\,cm 
diameter and are detected by a large--area two--dimensional position--sensitive 
detector built up from an array of six Charge--Coupled devices (CCDs)\,\cite{Nel02}.  
The hydrogen density is varied by temperature in a cryogenic target.

For the precision study of the $\pi H$ X--ray  line shapes the knowledge 
of the crystal spectrometer response is required at a higher level than available 
up--to--now. Narrow fluorescence X--rays or few keV $\Gamma$  lines for testing 
Bragg crystals are not available in practical cases. Therefore, the crystal response had been 
obtained from narrow pionic--atom transitions in previous 
experiments\,\cite{Sch01,Ana01}. However, the limited rates even at 
high--flux pion channels lead to unacceptable long measuring times for precision studies. 
For that reason, an Electron--Cyclotron--Resonance Ion Trap (ECRIT) source 
is being set up to produce hydrogen-- and helium--like electronic atoms. In such
few--electron atoms, fluorescence lines are narrow and can be used for thorough 
high--statistics studies of Bragg crystals\,\cite{Bir02}.

\section{First results}

A first series of measurements took place at the high--intensity 
pion channel $\pi$E5 of PSI. At a target density equivalent to  
3.5\,bar and using a H$_{2}$/O$_{2}$ gas mixture the $\pi$O(6h--5g) calibration 
line and the $\pi$H(3p--1s) transition were measured simultaneously 
(Fig\,2). This calibration method is basically free of systematic 
errors due to long--term instability. At higher densities hydrogen and oxygen have to 
be measured alternately to prevent the oxygen gas from freezing.

No density dependence was observed for the energy of the $3p-1s$ transition, 
which is interpreted as the absence of radiative decay during molecular formation. 
The energy values obtained are consistent within the errors. The weighted average
for the hadronic shift reads $\epsilon_{1s} = 7.120\pm 0.008 {~+\,0.009 \atop ~-\,0.008}\,~eV$.
The first error represents the statistical accuracy. The second one includes systematic 
effects, which are due to spectrometer setup, imaging properties of extended Bragg crystals, 
analysis and instabilities.

\begin{figure}[htb]
\begin{center}
\epsfig{file=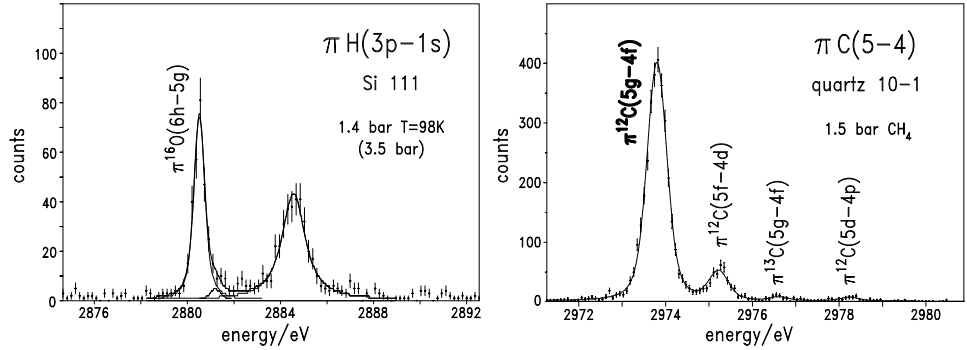,scale=0.67}
\caption{Left: Ground--state transition $3p-1s$ in pionic hydrogen measured
         simultaneously with the calibration transition $\pi^{16}O(6h-5g)$.
         Right: Pionic carbon $5-4$ transitions measured with a quartz crystal
         used to determine the spectrometer response function. A 
         resolution of about 450~meV was achieved.}
\end{center}
\label{fig:pih3p1s_piC}
\end{figure}

The measured line shape of the pionic hydrogen K transitions is a 
convolution of a Lorentz profile according to the natural width 
$\Gamma_{1s}$, the resolution of the crystal spectro\-meter and in
general several contributions to the Doppler width caused by various $n\rightarrow n'$ 
Coulomb transitions. A significant increase of the total width was found for the $2p-1s$ line 
compared to the $3p-1s$ transition (after deconvolution of the spectrometer response), 
which is attributed to the higher energy release available for the acceleration of the pionic--hydrogen 
system\,\cite{Hen03}. This result is corroborated by a reduced line width of the 
$4p-1s$ transition. On the other hand, no evidence for an increase of the line width with 
density was found even in the liquid phase.
From the $3p-1s$ and $4p-1s$ transitions, a save upper limit of  can be extracted 
$\Gamma_{1s} < 0.850\,eV$, which is slightly below the result of\,\cite{Sch01}.
A more refined analysis is going on.

\section{Conclusions and Outlook}

The preliminary result from this experiment corroborates the value for $\epsilon _{1s}$  found 
earlier\,\cite{Sch01}. The previous value for the hadronic width $\Gamma_{1s}$, however, 
exceeds the upper limit derived from this data indicating that the contributions from Coulomb
de--excitation are substantially underestimated. 

The forthcoming steps include in a detailed investigation of  Bragg crystals by using the PSI ECRIT 
and of Coulomb de--excitation by measuring muonic hydrogen. The line width  in $\mu H$ -- after 
deconvolution of the spectrometer response -- will be interpreted in terms of a new dynamical 
cascade picture involving the velocity of the exotic atom and the results from recent 
calculations for the cross section of the various collision processes\,\cite{Jen02}.

\end{document}